\documentstyle[12pt]{article}
\input psfig
\long\def\comment#1{}
\begin{document}
\title{Statistical Constraints on State
Preparation for a Quantum Computer}
\author{Subhash Kak\\
Department of Electrical \& Computer Engineering\\
Louisiana State University\\
Baton Rouge, LA 70803-5901, USA}
\date{March 26, 2001}
\maketitle

\begin{abstract}
Quantum computing algorithms require that the
quantum register  be initially present in a
superposition state.
To achieve this, we consider the practical problem of
creating a coherent superposition state of
several qubits.
We show that the constraints of quantum statistics
require that the entropy of the system 
be brought down
when several independent qubits are assembled together.
In particular, we have:
(i) not all initial states are realizable as pure states; 
(ii) the temperature of the system must be reduced.
These factors, in 
addition to decoherence and sensitivity to errors, must
be considered in
the implementation of quantum computers.\\

\vspace{2mm}
\noindent
{\bf Keywords.} Quantum computing; initializing a quantum state

\vspace{2mm}
\noindent
{\bf PACS No. 03.65.Bz}

\end{abstract}

\newpage

\section{Introduction}

There is a duality between the classical and the quantum:
In the classical world objects are distinct, while in the
quantum world they are indistinguishable. We, the users
of quantum computing, belong to the classical world.
This creates a conundrum for starting a computation on
a quantum computer. How do we load information on the quantum
register if the information-carrying particles in the cells
of the register are indistinguishable? 

Quantum computing algorithms as visualized now [1,2] proceed
with the register of $n$ cells in a pure state.
Each cell is seen to store a qubit
 $\alpha e^{i \theta_1} | 0\rangle + \beta  e^{i \theta
_2}| 1\rangle,$
where $ \alpha, \beta$ are real numbers and $\alpha^2 + \beta^2 = 1$.
Normally, the state of the register, $|\phi\rangle$, is taken to be the
all-zero state
of $n$-qubits: $| 0\rangle| 0\rangle ...| 0\rangle,$
or the amplitude vector $(1, 0, 0... 0)$, which,
by a process of rotation transformations 
on each qubit, is transformed into the 
state with amplitudes $(\frac{1}{\sqrt N},\frac{1}{\sqrt N},\frac{1}{\sqrt N}
... \frac{1}{\sqrt N}),$ where $N = 2^n $. In the
most general case, the state function 
can be written out as:

\begin{equation}
|\phi\rangle = \sum_{x=00...0}^{11...1} c_{x} | x\rangle 
\end{equation}
where the $c_x$ are complex numbers 
($\sum |c_x |^2 = 1 $) and the index $x$ ranges
over all $2^n$ values of an $n$-bit string. Quantum computing
is the application of appropriate unitary transformations on an 
initial state function that describes the problem to be
computed.

Implementation issues
related to decoherence and sensitivity to errors, after the
computation has started, have been considered in the literature [3].
In particular, several groups [4] have 
used trapped ion and NMR techniques to run simple quantum
algorithms.
But they have not demonstrated the initialization of
an arbitrary pure state in the register. It has been
shown [5], that the initial state 
loaded in these examples 
was a mixture, and so these early efforts
do not fully implement eqn. (1).
The NMR experiments are thus to be properly seen as
simulations of quantum computations rather than
true quantum computations.
 
There
exist several interesting problems with the model of eqn. (1)
regarding manipulations of the contents of a quantum register.
The
assumption inherent in the
algorithms that the phase uncertainty related
to the $c_x$s in each
of these superposition states is identical and it can be lumped
together and ignored has
been
questioned [6]. 
It has also been shown that it is impossible to delete an
unknown quantum state. In particular, if there are several
copies of an unknown photon, it is not possible to delete
the information content of one or more of these photons
by a physical process [7].
Likewise, it is not always possible to go from one state
to another using local transformations [8].
These problems arise due to the nature of
superpositions in a quantum state.

The problem related to the difficulty 
of the transformation of one quantum state to
another may be posed in other forms as well.
Here we consider the question of starting from
a suitable initial state on the register to reach the
amplitude vector
 $(\frac{1}{\sqrt N},\frac{1}{\sqrt N},\frac{1}{\sqrt N}
... \frac{1}{\sqrt N}),$
a superposition of all the components.
We show that statistical 
constraints make it impossible to do so
under ordinary circumstances.
However, we do not argue that it is impossible, 
in principle, to prepare such a state.

\section{Preparing the superposition state}

Consider a quantum register with $n$ cells,
each containing a qubit.
If the qubits are independent quantum systems,
and they are brought together, we
will have a mixture.
We can see this clearly
by imagining that the individual qubits are
physically remote from each other. 

The challenge is to obtain a superposition 
state that is coherent so that it can be
considered a single quantum system.
Qubits, generated separately and assembled together, do
not create the
appropriate superposition,
because this 
assembly ignores the constraints of
quantum statistics.
The process of bringing the qubits together
alters the conditions related to the
preparation of the qubits.
So how do the distinguishable particles which
are the starting qubits make a
transition to the superposition state of the
collection?

Before, answering this question, one must note that
the use of
classical notions in considering the contents of
the register can
lead to erroneous conclusions.
For a quantum system, it is essential to speak not
in terms of {\it a priori} properties, but in terms
of state preparation and observation.
Quantum mechanics is not a theory about reality;
it is a prescription for making the best possible
predictions about the future, if we have certain
information about the past [9].

A pure state is one which yields a specific outcome in 
a given test designed to elicit the maximum number
of outcomes associated with the system [6].
Examples are Stern-Gerlach experiment for spin or
the use of a calcite crystal for photon polarization.
One may represent the all-zero state by
$(1, 0, 0...0),$ if it is taken that each of the
qubits has been prepared identically and there is
no dynamical evolution. This is equivalent to considering that
qubits emerging out of the state preparation
apparatus are frozen in their state and installed
at the appropriate locations in the quantum register.

If the particles are generated at a certain point and
tested to yield a specific outcome and, if they
pass the test, transported to the right locations on the
register, there is no way to guarantee that each
of the particles would not have suffered dynamic evolution
prior to installation. If instead, $n$ tests are performed
simultaneously on the particles so as to cut down
on the delay, it may happen that some of these
tests do not yield the specified outcome and so
the failed tests will have to be repeated 
resulting in delay.

It appears more reasonable to assume that the qubits
are already available at the cell-sites of the
quantum register. These qubits then will be
individually steered to a specific pure state.
This assumption is actually made in the
description of the process to obtain the
superposition state required at the
start of the quantum algorithm.

\section{Collection of qubits}

But, having done this, it is necessary to
examine the collection of the $n$ qubits
from a statistical perspective.
As a quantum system, the register cannot
be viewed as consisting of unique particles,
which is what is implicit in the
individual rotation of the qubits to
obtain the superposition state of
$2^n$ components.
We must remember that not every pure state is realizable.

Only classical particles in a $n$-cell
register, each with
two states, can be distinguished
in terms of $2^n$ total components. 
Quantum particles are indistinguishable
and this restricts the number of possible
distinct states.
There will be the usual symmetry restrictions
associated with the state function of the register
depending on the particles being bosons or fermions.

We wish to stress the issue of indistinguishability
of particles in a superposition state.
For example, a $n$-cell register
with polarized photons in each cell
can have only $n+1$ distinguishable states.
If only one of the photons out of $n$ is polarized
vertically, then this particular photon is
not to be localized to a specific cell of
the register.
Although the measurement apparatus will localize
the vertically polarized photon at one of the
$n$-locations, this vertical polarization will
be shared by all the $n$-cells and so its
appearance at a specific location must be viewed
in a probabilistic sense as a part of the
state reduction process.

To see this further, consider a register of 3-cells
where one of the three qubits is in the state 
$| 1 \rangle$ and the other two are in the state
$| 0 \rangle$. Since the particles are indistinguishable,
it is incorrect to write the state of the register as
$| 100\rangle$,
where it is assumed that the first cell has a
$|1\rangle$ qubit and the other two have
$|0\rangle$ qubits. 
Just like an electron cannot be localized in a box
before it has been observed, a specific quantum
property cannot be localized to a particle that belongs
to a collection. 
The correct state
description for this case is:

\begin{equation}
 |\phi\rangle = \frac{1}{\sqrt 3} |100\rangle +
 \frac{1}{\sqrt 3} |010\rangle + 
 \frac{1}{\sqrt 3} |001\rangle
\end{equation}
where the relative phases have been ignored.

A quantum register will still yield
$n$-bits of information.
But the indistinguishability of the particles
throws a veil over the quantum reality which
limits our capacity to structure the states
on the register and to speak of a specific
characteristic of the particles in the set.

It is indeed true that the atoms at different
locations, carrying qubits, are 
physically distinct. 
But such
distinct qubits are normally in 
a separable, mixed state.
In a mixed state, the quantum
properties are indeed localized, but such a state cannot
be used to start the computation on a
quantum computer.

\section{Conclusions}

The representation of the state of the register
in terms of $2^n$ components, which is the
starting point of most quantum computing algorithms, is
contingent on steering an initial 
state, which appears to be unrealizable.
If the starting state cannot be realized,
then such quantum computing models can only
be taken to be 
mathematical constructs, not in accord with physical reality.

If the cells of the register are independent
quantum systems, then the uniqueness of the
contents can be maintained. But in this case,
the cells must be coupled for any useful
computation to be possible. 
The couplings and the resultant entanglements
form complications in the model beyond the
scope of the standard quantum computing paradigm.
This scenario departs from the
usual one and the physical constraints
necessary to be satisfied for this system
to process superpositions of qubits need
to be examined.

The reduction in the number of distinguishable states from
$2^n$ components of the $n$ qubits to the $n+1$ distinguishable
states of the coherent quantum system means that there
is a corresponding reduction in entropy from $n$
to $log_2 (n+1)$.
This means that energy equal to 
\begin{equation}
[n - log_2 (n+1) ] k T ln 2
\end{equation}
must be removed from the system [10] for
a computation to proceed. But, as explained earlier, this computation
cannot be based on {\it a priori} assignment of
states to the individual qubits. 

This analysis indicates that quantum statistical constraints need to be
considered in the formulation of quantum algorithms and
in their implementation. These constraints add to the difficulty
of initialization of the state on the quantum register before the
start of the computation.

We do not claim that these constraints will apply to all
quantum computational schemes. This makes it important
to look for such computational schemes where this is
not a problem. Likewise, one needs to look for effective
ways of preparing certain states of $n$-qubits.

\newpage
\subsection*{References}

\begin{enumerate}

\item
D. Deutsch, {\it Proc. R. Soc. Lond. A} 425, 73 (1989).

\item
P.W. Shor,
{\it SIAM J. Computing} 26, 1474 (1997),\\
L.K. Grover, 
{\it Physical Review Letters}
79, 325 (1997).

\item 
A. Ekert and R. Jozsa, {\it Rev. of Mod. Phys.} 68, 733 (1996),\\
S. Bose et al, {\it Phil.Trans.Roy.Soc.Lond.} A356, 1823 (1998).

\item 
N. Gershenfeld and I.L. Chuang, {\it Science} 275, 350 (1997),\\
D.G. Cory et al, {\it Proc. Natl. Acad. Science USA}  94, 1634 (1997),\\
J.A. Jones et al, {\it Nature} 403, 869 (2000),\\
L.M.K. Vandersypen et al, {\it Phys. Rev. Lett.} 85, 5452 (2000).

\item 
S.L. Braunstein et al, {\it Phys. Rev. Lett.} 83, 1054 (1999).

\item
S. Kak, 
{\it Foundations of Physics} 29, 267 (1999),\\
S. Kak, 
{\it Information Sciences} 128, 149 (2000).

\item
A.K. Pati and S.L. Braunstein, {\it Nature} 400, 164 (2000).

\item 
D. Jonathan and M.B. Plenio, {\it Phys. Rev. Lett.} 83, 3566 (1999).

\item
A. Peres, {\it Quantum Theory: Concepts and Methods}.
Kluwer Academic, Dordrecht, 1995.

\item
S. Kak, 
{\it Foundations of Physics} 26, 1375 (1996).

\end{enumerate}
 
\end{document}